\documentclass[preprint,authoryear]{elsarticle}
\usepackage{fullpage}
\usepackage{subcaption}
\usepackage{pifont}
\usepackage{latexsym}
\usepackage{mathrsfs}
\usepackage{amssymb,amsbsy}
\usepackage{amsfonts}
\usepackage{graphicx}
\usepackage{tcolorbox}
\usepackage[colorinlistoftodos]{todonotes}
\usepackage{varwidth}
\usepackage{algorithm}
\usepackage{algorithmic}
\usepackage{lmodern}
\usepackage{natbib}
\usepackage{placeins}
\usepackage{mathtools}      
\usepackage[titletoc]{appendix}
\usepackage{titletoc}
\usepackage{ntheorem}
\usepackage{float}
\usepackage{multirow}

\setcitestyle{authoryear}
\def\berr{\begin{eqnarray}}
\def\err{\end{eqnarray}}
\def\berrno{\begin{eqnarray*}}
\def\errno{\end{eqnarray*}}
\def\be{\begin{equation}}
\def\ee{\end{equation}}

\journal{Astronomy and Computing}

\begin{document}
\bibliographystyle{plainnat}
\begin{frontmatter}

\title{Theoretical Validation of Potential Habitability via Analytical and Boosted Tree Methods: An Optimistic Study on Recently Discovered Exoplanets}

\author[1]{Snehanshu Saha}
\author[1]{Suryoday Basak}
\author[2]{Kakoli Bora\corref{cor1}}
\ead{k\_bora@pes.edu}
\author[3]{Margarita Safonova}
\author[1]{Surbhi Agrawal}
\author[1]{Poulami Sarkar}
\author[4]{Jayant Murthy}

\cortext[cor1]{Corresponding author}

\address[1]{Department of Computer Science and Engineering, PESIT-BSC, Bangalore}
\address[2]{Department of Information Science and Engineering, PESIT-BSC, Bangalore}
\address[3]{M.~P.~Birla Institute of Fundamental Research, Bangalore}
\address[4]{Indian Institute of Astrophysics, Bangalore}

\begin{abstract}
Seven Earth-sized planets, known as the TRAPPIST-1 system was discovered with great fanfare in the last week of February 2017. Three of these planets are in the habitable zone of their star, making them potentially habitable planets a mere 40 light years away. Discovery of the closest potentially habitable planet to us just a year before -- Proxima~b and a realization that Earth-type planets in circumstellar habitable zones are a common occurrence provides the impetus to the existing pursuit for life outside the Solar System. The search for life has two goals essentially: Earth similarity and habitability. An index was recently proposed, Cobb-Douglas Habitability Score (CDHS), based on Cobb-Douglas habitability production function, which computes the habitability score by using measured and estimated planetary parameters like radius, density, escape velocity and surface temperature of a planet. The proposed metric, with exponents accounting for metric elasticity, is endowed with analytical properties that ensure global optima and can be scaled to accommodate a finite number of input parameters. We show here that the model is elastic, and the conditions on elasticity to ensure global maxima can scale as the number of predictor parameters increase. K-Nearest Neighbor classification algorithm, embellished with probabilistic herding and thresholding restriction, utilizes CDHS scores and labels exoplanets to appropriate classes via feature-learning methods. The algorithm works on top of a decision-theoretical model using the power of convex optimization and machine learning. The goal is to classify the recently discovered exoplanets into the ``Earth League" and other classes. A second approach, based on a novel feature-learning and tree-building method classifies the same planets without computing the CDHS of the planets and produces a similar outcome. The convergence of the two different approaches indicates the strength of the proposed scheme and the likelihood of the potential habitability of the recent discoveries.
\end{abstract}

%\emph{\textbf{Keywords}:} Habitability Index, Cobb Douglas Production Function, Exoplanets, Machine Learning, Cobb-Douglas Habitability Score, Optimization.

\end{frontmatter}

\section{Introduction}\label{sec:intro}

With discoveries of exoplanets pouring in hundreds, it is becoming necessary to develop some sort of a quick screening tool -- a ranking scale -- for evaluating habitability perspectives for the follow-up targets. We have proposed a novel inductive approach, inspired by the Cobb-Douglas model from production economics, to verify theoretical conditions of global optima of the functional form to model and to compute the habitability score of exoplanets -- the Cobb-Douglas Habitability Score (CDHS) \citep{Bora2016CDHS}. While our paper ``CD-HPF: New Habitability Score Via Data Analytic Modeling" was in production, the discovery of an exoplanet Proxima~b orbiting the nearest star (Proxima Centauri) to the Sun was announced \citep{Anglada-Escudé}. This planet generated a lot of stir in the news \citep{Witze} because it is located in the habitable zone and its mass is in the Earth's mass range: $1.27 - 3$ M$_{\oplus}$, making it a potentially habitable planet (PHP) and an immediate destination for the Breakthrough Starshot initiative \citep{Starshot}. A few months after the announcement of Proxima b, another family of terrestrial-size exoplanets -- the TRAPPIST-1 system -- was discovered \citep{Gillon}.

 This work is motivated by testing the efficacy of the suggested model, CDHS, in determining the habitability score, the proximity to the ``Earth-League", of the recently discovered Proxima~b. The habitability score model has been found to work well in classifying previously known exoplanets in terms of potential habitability. Therefore it was natural to test whether the model can also classify it as potentially habitable by computing its habitability score. This could indicate whether the model may be extended for a quick check of the potential habitability of newly discovered exoplanets in general. As we see in \textbf{Section~\ref{sec:trappist}}, this is indeed the case with the TRAPPIST-1 planets.  

\begin{figure*}
\centering
\includegraphics[width=1\columnwidth]{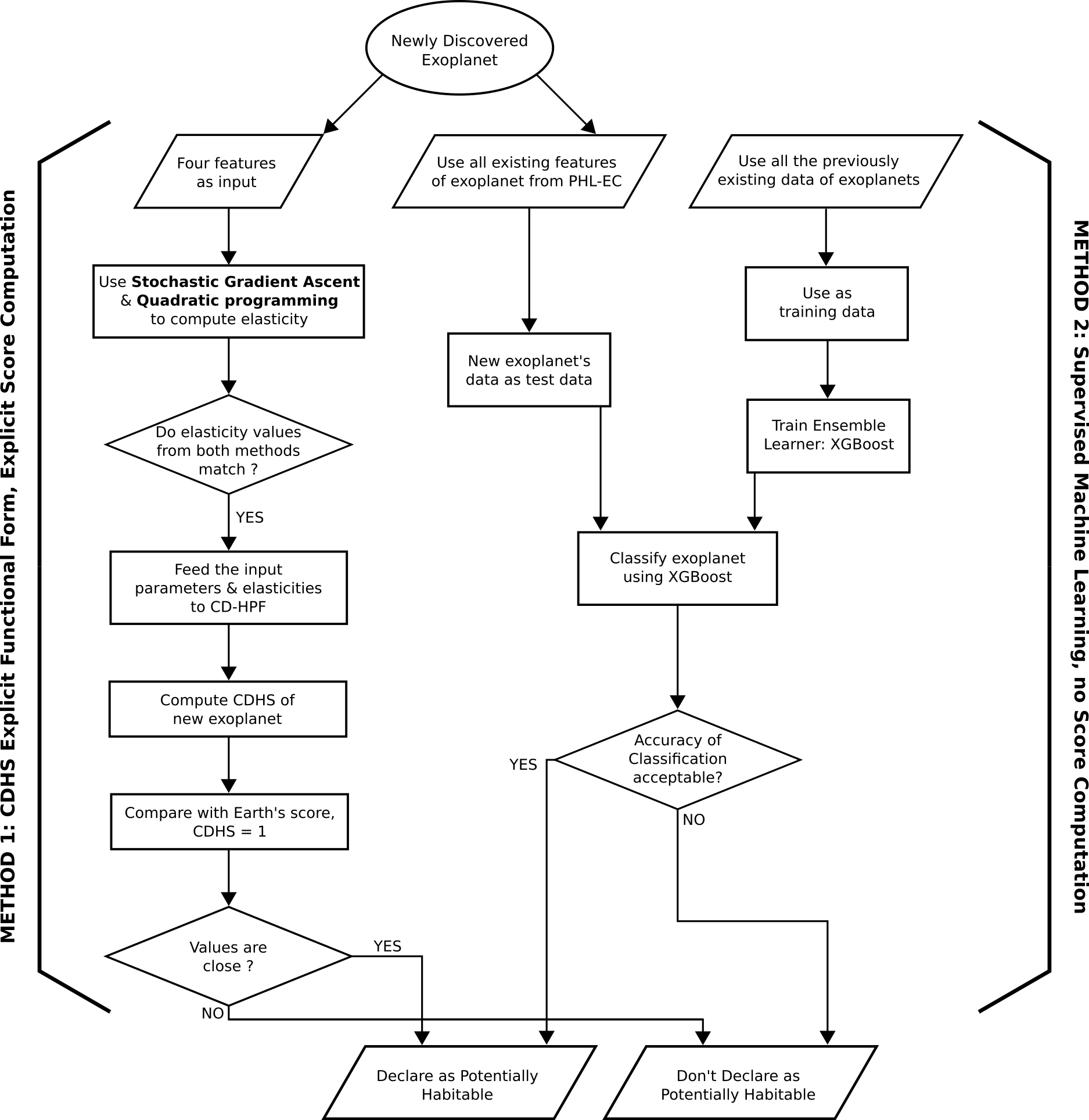}
\caption{The convergence of two different approaches in the investigation of potential habitability. The outcome of the explicit scoring scheme placed Proxima~b in the ``Earth League", which is synonymous to being classified as potentially habitable.}
\label{fig:flowchart}
\end{figure*} 

The flowchart in Figure~\ref{fig:flowchart} summarizes our new approach to the habitability investigation of exoplanets.
(on the example of Proxima~b and TRAPPIST-1 system). 
This approach is based on combination of two methods. The outcome of classification of exoplanets based on the CDHS (Method~1) is tallied with another classification method which discriminates samples (exoplanets) into classes based on the features/attributes of the samples (Method~2). The similar outcome from both approaches (the exoplanets are classified into the same habitability class), markedly different in structure and methodology, fortifies the growing advocacy of using machine learning in astronomy. 

 The habitability score model considers four parameters/features, namely mass, radius, density and surface temperature of a planet extracted from the PHL-EC (Exoplanet Catalog hosted by the Planetary Habitability Laboratory (PHL))\footnote{The latest updated (May 2017) dataset can be downloaded from the PHL website: http://phl.upr.edu/projects/habitable-exoplanets-catalog/data/database}. Though the catalog contains 68 observed and derived stellar and planetary parameters, we have currently considered only four for the CDHS model. CDHS does encounter problems commonly found in convex functional modeling, such as scalability and curvature violation. We show here that the CDHS model is scalable, i.e. capable of accommodating more parameters (see Section \ref{sec:model_scalability} on model scalability, and Section 3 in supplementary file (Proof of model scalability), \cite{Supp_File} for the proof of the theorem). Therefore, we may use more parameters in future to compute the CDHS. The problem of curvature violation is tackled in Sec.\ref{sec:model_scalability} later in the paper. 

PHL classifies all discovered exoplanets into five categories based on their thermal characteristics: non-habitable, and potentially habitable: psychroplanet, mesoplanet, thermoplanet, and hypopsychroplanet. Proxima~b and the TRAPPIST-1 system are amongst the recent additions to the catalog with recorded features. Here, we employ a non-metric classifier to predict the class label of the recently discovered exoplanets. We compute the accuracy of our classification method, and aim to reconcile the result with the habitability score of the recently discovered exoplanets, which may suggest its proximity to the ``Earth League". We call this an investigation in the optimistic determination of potential habitability. The hypothesis is the following: a machine learning-based classification method, known as boosted trees, classifies exoplanets and returns some with the class by mining the features present in the PHL-EC (Method~2 in Figure~\ref{fig:flowchart}). This process is independent of computing the explicit habitability score for recently developed exoplanets ({\em aka} Method~1 in Figure~\ref{fig:flowchart}), and indicates habitability class by learning attributes from the catalog. This implicit method should match the outcome suggested by the CDHS. In other words, the class label of exoplanets predicted by the implicit method (Method 2) should correspond to the appropriate CDHS of those exoplanets. This is demonstrated in Table~\ref{t:aug:abhijit}. 

The second approach is based on XGBoost -- a statistical machine-learning classification method used for supervised learning problems, where the training data with multiple features are used to predict a target variable. Authors intend to test whether the two different approaches to investigate the habitability of Proxima~b and the TRAPPIST-1 planets, analytical and statistical, converge with a reasonable degree of confidence. The paper is split into two layers. The first layer considers Proxima~b as a test case, and the latter layer applies our methods on the Trappist-1 system. Trappist-1 was discovered later than Proxima~b and, hence the section  Sec.~\ref{sec:trappist} was added.

The paper is organized as follows.
Sections~\ref{sec:analytical_cdhs},~\ref{sec:sga}, ~\ref{sec:model_scalability}, and \ref{sec:prox_xgb} elaborate the theory and methods and discuss the implications on Proxima~b as a test case (Layer 1). 
Section~\ref{sec:trappist} forms layer 2, focusing on the Trappist-1 system. Section~\ref{sec:conclusion} elaborates the overall efficacy of our approach applied to Proxima~b and Trappist-1 system. Supplementary file \citep{Supp_File} presents the theory of CDHS (sections 3, 4 and 5) and classification by boosted trees (section 6) in detail.

\section{Analytical Approach via CDHS} \label{sec:analytical_cdhs}

We begin by discussing the key elements of the analytical approach. The parameters of Proxima~b were extracted from the PHL-EC:
minimal mass 1.27 EU, radius 1.12 EU, density 0.9 EU, surface temperature 262.1 K, and escape velocity 1.06 EU, where EU is the Earth Units. Its Earth Similarity Index (ESI), estimated using a simplified version\footnote{http://phl.upr.edu/projects/earth-similarity-index-esi} of the ESI, is 0.87. By definition, ESI range is from 0 (totally dissimilar to Earth) to 1 (identical to Earth), only planets with ESI $\geq 0.8$ are  considered to be Earth-like. 

\subsection{Earth Similarity Index}

In general, the ESI value of any exoplanet's planetary property is calculated using the following expression \citep{Schulze2011Two-tired}, 
\begin{equation}
ESI_{x}= \left( 1-\left | \frac{x-x_{0}}{x+x_{0}} \right | \right)^{w_{x}}\,,
\label{eq:ESI}
\end{equation}
where $x$ is a planetary property -- radius, surface temperature, density, or escape velocity, $x_0$ is the Earth's reference value for that parameter, i.e. 1 EU, 288 K, 1 EU and 1 EU, respectively, and $w_x$ is the weighted exponent for that parameter. After calculating ESI for each parameter by Eq.~\ref{eq:ESI}, the global ESI is found by taking the geometric mean (G.M.) of all four ESI$_x$,  
\begin{equation}
ESI=\left(\prod_{x=1}^n ESI_{x}\right)^{\frac{1}{n}} \,.
\label{eq:globalESI}
\end{equation} 
The problem in using Eq.~\ref{eq:globalESI} to obtain the global ESI is that sometimes there no available data to obtain all input  parameters, such as in the case of Proxima~b -- only its mass and the distance from the star are known. Due to that, a simplified expression was proposed by the PHL for ESI calculation in terms of only radius and stellar flux, 
\begin{equation}
ESI= 1-\sqrt{\frac{1}{2}\left(\frac{R-R_0}{R+R_0}\right)^2 + \left(\frac{S-S_0}{S+S_0}\right)^2}\,,
\label{eq:simpleESI}
\end{equation}
where $R$ and $S$ represent radius and stellar flux of a planet, and $R_0$ and $S_0$ are the reference values for the Earth. Using 1.12 EU for the radius and 0.700522 EU for the stellar flux, we obtain ESI = 0.8692. It is worth mentioning that once we know one observable -- the mass --  other planetary parameters used in the ESI computation (radius, density and escape velocity) can be calculated based on certain assumptions. For example, the small mass of Proxima~b suggests a rocky composition. However, since 1.27 EU is only a low limit on mass, it is still possible that its radius exceeds 1.5 -- 1.6 EU, which would make Proxima~b, not rocky \citep{rogers}. In the PHL-EC, its radius is estimated using the mass-radius relationship

\begin{equation}
 R=
\begin{cases}
M^{0.3}&M\leq 1\\
M^{0.5}&1\leq  M<200\\
\left(22.6\right)M^{(-0.0886)}&M\geq 200
\end{cases}
\label{eq:mass-radius}
\end{equation}
Since Proxima~b mass is 1.27 EU, the radius is $R=M^{0.5} \equiv 1.12$ EU. Accordingly, the escape velocity was calculated by $V_e=\sqrt{2GM/R}\equiv 1.065$ (EU), and the density by the usual $D=3M/4\pi R^3\equiv 0.904$ (EU) formula. If we use all four parameters provided in the catalog, the global ESI becomes 0.9088. 

\subsection{Cobb Douglas Habitability Score (CDHS)}
\label{sec:CDHS}

We have proposed the new model of the habitability score in \citep{Bora2016CDHS} using a convex optimization approach \citep{Saha2016optimization}. In this model, the Cobb Douglas function \citep{cobb-douglas} is reformulated as Cobb-Douglas habitability production function (CD-HPF) to compute the habitability score of an exoplanet,
\begin{equation}
\mathbb{Y}=f\left(R,D,T_{s},V_{e}\right)=K \left(R\right)^{\alpha}\cdot \left(D\right)^{\beta}
\cdot \left(T_{s}\right)^{\gamma}\cdot\left(V_{e}\right)^{\delta}\,
\label{eq:CDHS}
\end{equation}  
where the same planetary parameters are used -- radius $R$, density $D$, surface temperature $T_s$, and escape velocity $V_e$. $\mathbb{Y}$ is the habitability score CDHS, and $f$ is defined as CD-HPF \footnote{Elasticities $K$, $\alpha$, $\beta$, $\gamma$ and $\delta$ need to be estimated}. The goal is to maximize the score, $\mathbb{Y}$, where the elasticity values of each parameter are subject to the condition $\alpha+\beta+\gamma+\delta<1$. Note that the interior CDHS$_i$, denoted by $Y1$, is calculated using radius $R$ and density $D$, while the surface CDHS$_s$, denoted by $Y2$, is calculated using surface temperature $T_s$ and escape velocity $V_e$. The objective is to find elasticity value that produces the optimal habitability score for the exoplanet, i.e. to find
$Y_1 = \max_{\alpha,\beta} Y(R,D)$ such that,
$\alpha > 0$, $\beta>0$ and $\alpha+\beta\leq 1$. Similarly, we need to find $Y_2 = \max_{\gamma,\delta} Y(T,V_e)$ such that $\gamma > 0$, $\delta>0$ and $\delta+\gamma\leq 1$. Elasticity values are obtained by a computationally fast Stochastic Gradient Ascent (SGA) algorithm described in Sec.~\ref{sec:SGA}. We calculate CDHS score for the constraints known as returns to scale: Constant Return to Scale (CRS) and Decreasing Return to Scale (DRS) (for details, refer \citep{Bora2016CDHS}). Note that $\alpha+\beta < 1$ is the DRS condition for elasticity, which may be scaled to $\alpha_1+\alpha_2+\ldots+\alpha_n < 1$. Analogously, $\delta+\gamma < 1$ is the DRS condition for elasticity which may be scaled to $\delta_1+\delta_2+ \ldots + \delta_n < 1$.

As Proxima~b is considered an Earth-like planet, we endeavored to cross-match the observation via the method explained in the previous section. The analysis of CDHS will help to explore how this method can be effectively used for  newly discovered planets. The eventual classification of any exoplanet is accomplished by using the proximity of CDHS of that planet to the Earth, with additional constraints imposed on the algorithm termed ``probabilistic herding". The algorithm works by taking a set of values in the neighborhood of 1 (CDHS of Earth). A threshold of 1 implies that CDHS value between 1 and 2 is acceptable for membership in the ``Earth-League", pending fulfillment of further conditions. For example, the CDHS of the most potentially habitable planet before Proxima~b, Kepler-186~f, is 1.086 (the closest to the Earth's value), though its ESI is only 0.64. While another PHP -- GJ-163~c has the farthest score (1.754) from 1; and though its ESI is 0.72, it may not be even a rocky planet as its radius can be between 1.8 to 2.4 EU, which is not good for a rocky composition theory (see e.g. \citep{rogers}). 

Sometimes, values of certain parameters are not available in the catalog (e.g. for 11 planets PHL-EC does not provide surface temperatures). In machine learning, the missing values can be imputed by using association rules, in particular, the \textit{rule-based learning}. We have devised an algorithm based on \citep{Agrawal1993} and \citep{Agrawal1994} to impute missing values, the details of which are explained in supplementary file (section 1 of \citep{Supp_File}).

\subsection{CDHS calculation using radius, density, escape velocity and surface temperature}
Using the values of the parameters from the PHL-EC, we calculated CDHS score for the CRS and DRS cases, and obtained optimal elasticity and maximum CDHS value. The CDHS values in CRS and DRS cases were 1.083 and 1.095, respectively. The degree/extent of closeness is explained in \citep{Bora2016CDHS} in great detail. 

\subsection{CDHS calculation using stellar flux and radius}

Following the simplified version of the ESI (Eq.~\ref{eq:simpleESI}), we repeated the CDHS computation using only radius and stellar flux (1.12 EU and 0.700522 EU, respectively). From the scaled down version of Eq.~\ref{eq:CDHS}, we obtain CDHS$_{CRS}$ and CDHS$_{DRS}$ as 1.083 and 1.095, respectively.
These values confirm the robustness of the method used to compute CDHS and validate the claim that Proxima~b falls into the ``Earth-League" category.

\subsection{CDHS calculation using stellar flux and mass}

The habitability score requires the use of available physical parameters, such as radius, or mass, and temperature, and the number of parameters is not extremely restrictive. As long as we have the measure of the interior similarity -- the extent to which a planet has a rocky interior, and exterior similarity -- the location in the HZ or the favorable range of surface temperatures, we can reduce (or increase) the number of parameters. Since radius is calculated from an observable parameter -- mass, we decided to use the mass directly in the calculation, obtaining CDHS$_{DRS}$ as 1.168 and CDHS$_{CRS}$ as 1.196. The CDHS achieved using radius and stellar flux (previous subsection) and the CDHS achieved using mass and stellar flux have the same values.

\textbf{\textit{Remark:}} \textit{Does this imply that stellar flux and planet mass are enough to compute the habitability score as defined by our model? It cannot be confirmed until enough number of clean data samples are obtained containing the four parameters used in the original ESI and CDHS formulation. We plan to perform a full-scale dimensionality analysis as future work} 

The values of ESI and CDHS using different methods are summarized in Table~\ref{tab:table1}.

\begin{table}[h!]
\caption{ESI and CDHS values calculated for different parameters}\label{tab:table1}
\centering
\begin{tabular}{c c c c } 
%inserts double horizontal lines
%[0.5ex]
\hline
Parameters Used & ESI & CDHS$_{CRS}$ & CDHS$_{DRS}$\\
%[0.5ex]
\hline
$R$, $D$, $T_s$, $V_e$ & 0.9088 & 1.083 & 1.095\\ 
%modeled $T_s$, $R$ & 0.9026 & 1.095 & 1.084\\ 
Stellar Flux, $R$ & 0.869 & 1.196 & 1.168\\ 
Stellar Flux, $M$ & 0.849 & 1.196 & 1.167\\
\hline
\end{tabular}
\end{table}

NOTE: The nicety in the result, i.e. little difference in the values of CDHS, is due to the flexibility of the functional form in the model proposed in  \citep{GindeEconometric}, and the computation of the elasticities by the Stochastic Gradient Ascent method described in the next section. Using this method led to the fast convergence of the elasticities. Proxima~b passed the scrutiny and is classified as a member of the ``Earth League".

\section{Elasticity computation: Stochastic Gradient Ascent (SGA)} \label{sec:sga}

\citep{Bora2016CDHS} used a library function {\bf fmincon} to compute the elasticity values. Here, we have implemented a more efficient algorithm to perform the same task. This was done for two reasons: to be able to break free from the in-built library functions, and to devise a sensitive method which would mitigate oscillatory nature of Newton-like methods around the local minima/maxima. There are many methods which use gradient search, including the one proposed by Isaak Newton. Although theoretically sound, algorithmic implementations of most of these methods face convergence issues in real time due to the oscillatory nature. 

We have employed a modified version of the descent, an SGA algorithm, to calculate the optimum CDHS and the elasticities for mass, radius, density and escape velocity (Eq.~\ref{eq:CDHS} in Sec.~\ref{sec:CDHS}). As opposed to the conventional Gradient Ascent/Descent method, where the gradient is computed only once, stochastic version recomputes the gradient for each iteration and updates the elasticity values. Theoretical convergence, guaranteed otherwise in the conventional method, is sometimes slow to achieve though. Stochastic variant of the method speeds up the convergence, justifying its use in the context of the problem (the size of data, i.e. the number of discovered exoplanets, is increasing every day).

Output elasticity ($\alpha$, $\beta$, $\gamma$ or $\delta$) of Cobb-Douglas habitability function is the accentual change in the output in response to a change in the levels any of the inputs. Accuracy in elasticity values is crucial in deciding the right combination for the optimal CDHS, where different approaches are analyzed before arriving at final decision. In the next subsections, we show how the elasticities were computed on the example of $\alpha$ and $\beta$. Once they are computed, we repeat the procedure to compute other elasticities, $\gamma$ and $\delta$.

\subsection{Computing Elasticities via Gradient Ascent}
\label{sec:SGA}

Gradient Ascent is an optimization algorithm used for finding the local maximum of a function. Given a scalar function $F(x)$, gradient ascent finds the $\max_x F(x)$ by following the slope of the function. This algorithm selects initial values for the parameter $x$ and iterates to find the new values of $x$ which maximizes $F(x)$ (here CDHS). Maximum of a function $F(x)$ is computed by iterating through the following step, 
\begin{equation}
x_{n+1}\leftarrow x_{n}+\chi \frac{\partial F}{\partial x}\,,
\end{equation}
where $x_n$ is an initial value of $x$, $x_{n+1}$ the new value of $x$, 
\(\frac{\partial F}{\partial x}\) is the  slope of function $Y = F(x)$ and
$\chi$ denotes the step size, which is greater than $0$ and forces the algorithm to make a small jump (descent or ascent algorithms are trained to make small jumps in the direction of the new update). Stochastic variant thus mitigates the oscillating nature of the global optima -- a frequent malaise in the conventional Gradient Ascent/Descent and Newton-like methods, such as {\bf fmincon} used in \citep{Bora2016CDHS}. At this point of time, without further evidence of recorded/measured parameters, it may not be prudent to scale up the CD-HPF model by including more parameters other than the ones used by either ESI or our model. But if it ever becomes a necessity (to utilize more than the four parameters), the algorithm will come in handy and multiple optimal elasticity values may be computed fairly easily.

\subsection{Computing Elasticities via Constrained Optimization}

Let the assumed parametric form be $\log(y) = \log(K) + \alpha \log(S) + \beta \log(P)$\footnote{This is a logarithmic transformation of the standard CDHS model (which has the exponential form).}. Consider a set of data points,
\begin{equation}
\begin{array}{ccccccc}
\ln(y_1) & = & K' &{}+{} & \alpha S'_1  & {}+{} & \beta P'_1  \\
 \vdots & & \vdots & & \vdots & & \vdots \\
\ln(y_N) & = & K' &{} +{} & \alpha S'_N & {}+{} & \beta P'_N  
\end{array}
\label{eq:system}
\end{equation}
where $K' = \log(K)\,, \ S'_i = \log(S'_i)$ and  $P'_i = \log(P'_i)$. If $N > 3$, this is an over-determined system, where one possibility to solve it is to apply a least squares method. Additionally, if there are constraints on the variables (the parameters to be solved for), this can be posed as a constrained optimization problem. These two cases are discussed below.\\
\textbf{No constraints:} This is an ordinary least squares solution. The system  is in the form $y = Ax$, where
\begin{equation}
x =
 {\begin{bmatrix}
    K' & \alpha & \beta
  \end{bmatrix}}^T\,,\quad y = 
\begin{bmatrix}
    y_1\\ . \\ . \\ y_N
  \end{bmatrix}\,,
  \label{eq:xy_matrices}
\end{equation}
and
\begin{equation}
A = \begin{bmatrix}
    1 & S'_1 & P'_1\\ 
      & ...  &     \\ 
    1 & S'_N & P'_N
  \end{bmatrix}\,.
\end{equation}

The least squares solution for $x$ is the solution that minimizes 
\begin{equation}
(y - Ax)^T (y - Ax)\,.
\end{equation}
It is well known that the least squares solution to Eq.~\eqref{eq:xy_matrices} is the solution to the system $A^T y= A^TAx $, i.e. $x = (A^TA)^{-1} A^T y$. In {\em Matlab}, the least squares solution to the overdetermined system $y = Ax$ can be obtained by
$x = A / y$. Table \ref{tab:irscrsdrs-noconstraints} presents the results of least squares ({\bf no constraints}) obtained for the elasticity values after performing the least square fitting, while Table \ref{tab:irscrsdrs-withconstraints} displays the results obtained for the elasticity values after performing the constrained least square fitting; in Table \ref{tab:irscrsdrs-quadprog}, the values of CRS and DRS from quadratic programming have been enunciated.

\begin{table}[!htbp]
\centering
\caption{Elasticity values for IRS, CRS \& DRS cases after performing the least square test ({\bf no constraints}): elasticities $\alpha$ and $\beta$ satisfy the theorem $\alpha + \beta < 1,  \alpha + \beta = 1$, and $\alpha + \beta > 1$ for DRS, CRS and IRS, respectively, and match the values reported previously in \cite{Bora2016CDHS}. }\label{tab:irscrsdrs-noconstraints}
\begin{tabular}{|c|c|c|c|}
\hline
& IRS & CRS & DRS\\
\hline
\(\alpha\) & 1.799998 & 0.900000 & 0.799998\\
\hline
\(\beta\) & 0.100001 &  0.100000 & 0.099999\\
\hline
\end{tabular}
\end{table}

\textbf{Constraints on parameters:} This results in a constrained optimization problem. The objective function to be minimized (maximized) is still the same, namely,
\begin{equation}
(y - Ax)^T (y - Ax)\,.
\end{equation}
This is a quadratic form in $x$. If the constraints are linear in $x$, then the resulting constrained optimization problem is a quadratic program (QP). A standard form of a QP is 
\begin{equation}
\max x^T H x +f^T x\,,
\label{eq:standard_form_constrained_optimization}
\end{equation}
such that
\berr
&Cx \leq b \,;\quad\,\,\,\,&\text{Inequality constraint}\nonumber\\
&C_{\rm eq}x = b_{\rm eq} \,;&\text{Equality constraint.}\nonumber
\err
Suppose the constraints are $\alpha, \beta >  0$ and $\alpha  + \beta \leq 1$. The QP can be written as (neglecting the constant term $y^Ty$)
\begin{equation}
\max x^T(A^TA)x - 2y^TAx\,,
\label{eq:17}
\end{equation}
such that
\begin{equation}\label{eq:18} 
\begin{cases}
\alpha > 0 \,, \\
\beta > 0 \,, \\
\alpha + \beta \le 1 \,.
\end{cases}
\end{equation}
For the standard form as given in Eq.~\eqref{eq:standard_form_constrained_optimization}, Eqs.~\eqref{eq:17} and \eqref{eq:18} can be represented by rewriting the objective function as:
\begin{equation}
x^THx + f^Tx\,,
\end{equation}
where
\begin{equation}
H = A^TA \text{  and  } f=-2A^Ty \,.
\end{equation}
The inequality constraints can be specified as
\begin{equation}
C =
 \begin{bmatrix}
    0 & -1 & 0\\
    0 &  0 & -1\\
    0 &  1 &  1 
  \end{bmatrix}\,,\ \text{and} \,\,
b =
 \begin{bmatrix}
    0 \\
    0 \\
    1 
  \end{bmatrix}\,.
\end{equation}

In {\em Matlab}, the QP can be solved using the function {\bf quadprog}. The results in Table \ref{tab:irscrsdrs-withconstraints} were obtained by conducting quadratic programming.
\FloatBarrier
\begin{table}[!htbp]
\centering
\caption{Elasticity values for IRS, CRS \& DRS cases after performing the least square test (\textbf{with constraints}): elasticity values $\alpha$ and $\beta$ satisfy the theorem $\alpha + \beta < 1,  \alpha + \beta = 1$, and $\alpha + \beta > 1$ for DRS, CRS and IRS, respectively, and match the values reported previously \citep{Bora2016CDHS}. } \label{tab:irscrsdrs-withconstraints}
\begin{tabular}{|c|c|c|c|}
\hline 
& IRS & CRS & DRS\\
\hline
\(\alpha\) & 1.799998 & 0.900000 & 0.799998\\
\hline
\(\beta\) & 0.100001 &  0.100000 & 0.099999\\
\hline
\end{tabular}
\end{table}

\textbf{Using active set:} We have conducted the experiment using active learning technique. This framework is best suited in our case as it can be applied to different performance targets and all types of classifications. The traditional active-set method is divided into two steps, focusing on feasibility and optimality, in that order. Instead of an ``ad-hoc" start,  active set methods bank on a good ``initiator" estimate of the optimal active set. This is well suited for a sequence of quadratic programs to be solved, which is what our constrained optimization problem needs. Active set gave best results out of all the three algorithms which suffices our argument.
\FloatBarrier
\begin{table}[!htbp]
\centering
\caption{Results of quadratic programming by using the active-set learning. Exact match with SGA results and Method 1, which satisfy the conditions of CRS, DRS \& IRS i.e. elasticity values $\alpha$ and $\beta$ satisfy the theorem $\alpha + \beta = 1; \alpha + \beta < 1; \alpha + \beta > 1$ and match the values reported in \citep{Bora2016CDHS}.} \label{tab:irscrsdrs-quadprog}
 \begin{tabular}{|c|c|c|}
\hline
 & CRS & DRS\\
 \hline
 K & 1 & 1\\
 \hline
 \(\alpha\)  & 0.9000 & 0.8000 \\
 \hline
 \(\beta\)  & 0.1000 & 0.1000\\
\hline
\end{tabular}
\end{table}

By solving the described constrained QP, we find that the result satisfy the condition $\alpha + \beta \leq 1$ for both CRS and DRS cases, and the condition $\alpha + \beta \geq 1$ for the IRS case. Elasticities $\alpha$, $\beta$ and $K$ from both  computations are very close, supporting our choice of $\alpha$, $\beta$ and $K$. Identical results are observed for elasticities $\gamma$ and $\delta$ for the surface CDHS. Obtained elasticity values used for computing CDHS are $\alpha=0.8$ and $\beta=0.1$ for DRS, and $\alpha=0.9$ and $\beta=0.1$ for CRS cases, respectively. The algorithms are repeated to compute $\gamma$ and $\delta$, and a convex combination of interior and surface CDHS is used to calculate the CDHS of Proxima~b. The entire process, including classification of all exoplanets post-habitability score computation, is summarized in Appendices.

\section{Model Scalability} \label{sec:model_scalability}

In paper \citep{Bora2016CDHS}, we have shown the theoretical guarantee regarding the conditions on elasticity. However, the scalability of the model (scalability of the theoretical guarantee) depends on the fact that the conditions of global maxima continue to hold even if the number of input parameters increase. In addition, the theoretical guarantee in some cases \citep{Saha2016optimization} tends to relax when an arbitrary parameter is added to the model. This happens due to curvature violation of the functional form. In other words, if eccentricity (say) is added as one of the input parameters along with surface temperature, density, radius and mass, there needs to be a mathematical guarantee that the conditions on elasticity should scale in the same fashion. This has been illustrated previously via computer simulation. However, there needs to be a theoretical result fortifying the intuitive understanding of the proposed model and the scoring scheme -- the CDHS. We define the  theorem which lays the foundation for model scalability in the event any parameter is added to the existing model, already accommodating an arbitrary number of parameters. If the conditions of elasticity for a global maxima hold for a fixed set of input parameters (say, $n$), it will continue to hold when the number of parameters is increased by 1 (say, $n + 1$). This is an inductive approach, non-traditional but powerful!  The proof (given in Section 3 of \citep{Supp_File}) is based on the principle of mathematical induction. \\

\textbf{Theorem:} If global maxima for CDHS, i.e.
\begin{equation}
\log(Y)=\frac{1}{1-\sum\limits_{i=1}^n \alpha_i}\log\left\{k\prod\limits_{i=1}^n \left(\frac{x_ip}{w_i}\right)^{\alpha_i}\right\} 
\end{equation}
holds, then the same condition for the global maxima will continue to hold if an additional input parameter is inserted in the habitability function CD-HPF, i.e., if
$$
\log(Y_{\rm new})=\frac{1}{1-\sum\limits_{i=1}^{n+1} \alpha_i}\log\left\{k\prod\limits_{i=1}^{n+1} \left(\frac{x_ip}{w_i}\right)^{\alpha_i}\right\}$$  
holds as well. Further, it follows that the elasticity condition for DRS for $n+1$ parameters is true, 
i.e. $ 1-\sum\limits_{i=1}^{m+1} \alpha_i > 0$, if the elasticity condition for DRS for $ n$ parameters, i.e. $ 1-\sum\limits_{i=1}^{m} \alpha_i > 0$, holds.\footnote{\underline{Remark}:The habitability score CDHS is computed using four parameters: $R$, $D$,$T_s$ and $V_e$. If a new parameter from the PHL-EC needs to be added to the CD-HPF, it is important to know if the conditions of global maxima for habitability still holds. The above theorem validates our superposition conclusively.}

We have investigated the habitability of newly discovered exoplanets via CDHS (Method~1). CDHS leads to a classification scheme (Algorithm~3 in Section 4 of the supplementary material, \citep{Supp_File}) and depends on computing the habitability score of discovered exoplanets. However, the classification problem doesn't have to rely on having numerical values of the response variable of samples under classification. Instead, the hidden relationship between samples may be discovered by construction of the decision rules connecting the feature values of the samples. In the next section, we will explore the habitability classification problem from a supervised learning perspective (Method~2), where a collection of labeled data (exoplanets from the PHL-EC) is used as training set, and Proxima~b is used as the test data. We train the machine to learn the features associated with the training and test data and identify the class label of the test data via machine classification algorithm, known as XGBoost. The goal, as stated earlier, is to test the ability of the algorithm to label Proxima~b in the ``Earth League" with a reasonably high accuracy, thereby establishing the strong correlation between the two different approaches.

\section{Classification of Proxima Centauri b via non-functional form: XGBoost, a feature-based learning and classification method} \label{sec:prox_xgb}

Here we illustrate a method by which the high habitability score of Proxima~b may be predicted by using class labels and features from the PHL-EC. The method XGBoost (eXtreme Gradient Boosting) is a non-metric classifier, and a fairly recent addition to the suite of machine learning algorithms \citep{Chen2016}. Non-metric classifiers are applied in scenarios where there are no definitive notions of similarity between feature vectors.

A typical machine-learning problem processes input data and combines that with the learning algorithm to produce a model as output. Learning implies recognizing complex patterns and making intelligent decisions based on data. The machine comes up with its own prediction rule,  based on which a previously unobserved sample would be classified as a certain type, meso or psychroplanets for example, with a reasonable accuracy.

In order to appropriately apply a method (including preprocessing and classification), a thorough study of the nature of the data should be done; this includes understanding the number of samples in each class, the separability of the data, etc. Depending on the nature of the data, appropriate preprocessing and post processing (if needed) methods should be determined along with the right kind of classifier for the task.

\subsection{Understanding the data to be classified}

The PHL-EC dataset contains more than 3500 samples and is growing steadily: from 1904 samples in November 2015 to 3635 samples at the time of writing \footnote{These numbers vary over time. We need these samples to train the classifier, so that it can classify new additions, such as e.g. Proxima~b or Trappist-1 planets.}. We have considered 51 features of the data for classifying, and have eliminated the ones that are unimportant for classification, such as the name of the parent star ({\it S.Name}), the name of the planet ({\it P.Name}), etc. In the dataset PHL-EC, planets are already segregated into five classes based on their surface thermal properties:
\begin{enumerate}
\item \textbf{Non-Habitable}: planets that do not have thermal properties required to sustain life.
\item \textbf{Mesoplanet}: planets with a mean global surface temperature between $0^\circ$C and $50^\circ$C -- a necessary condition for complex terrestrial life. These are generally referred to as Earth-like planets.
\item \textbf{Psychroplanet}: planets with mean global surface temperature between $-15^\circ$C and $+10^\circ$C -- somewhat colder than the optimal temperature for the sustenance of terrestrial life.
\item \textbf{Thermoplanet}: planets with the temperature in the range of $50^{\circ}$C -- $100^{\circ}$C -- warmer than the temperature range suited for most terrestrial life.
\item \textbf{Hypopsychroplanets}: planets with  temperature below $-50^{\circ}$C. These planets are too cold for the survival of most terrestrial life.
\end{enumerate}
Out of these, the classes of hypopsychroplanet and thermoplanet have too few samples (only two planets each) and hence are not useful for the analysis. The classification was performed on remaining three classes: psychroplanet, mesoplanet and non-habitable.

A planet having characteristics suitable for inhabitation is still a rare occurrence; naturally, most of the samples in the dataset belong to the class of non-habitable planets (3592 out of 3635). From a data analytic point of view, this is a \textit{data bias} and can lead to overfitting, i.e., when a classifier becomes overly complex and extremely sensitive to the nuances in the data. Overfitting is a problem that needs to be dealt with carefully and not be overlooked as an administrative task. In a dataset such as the PHL-EC, where the number of samples belonging to one class is over a thousand times the total number of samples belonging to all the other classes, just reporting the numeric accuracy obtained by directly feeding the data to train a classifier would be an incorrect methodology.

To counter the potential problems due to the dominance by a single class, we used \textit{artificially balanced} datasets by considering random samples from the classes of non-habitable and mesoplanets with the total number of samples belonging to one class being equal to the number of samples in the psychroplanet class, as it has the least number of samples. Then this balanced dataset was divided in a ratio 9:4 (we found this to be the best ratio), where the larger portion was that of the training set. This cycle of balancing the data set artificially, dividing it, training and testing a classifier was performed multiple times, and the mean accuracy of all the trials was considered to be representative of the potential of a classifier. By artificial balancing, the reported accuracies are also more reliable than without balancing.

We have applied a powerful ensemble classification for the task described above. Boosting refers to the method of combining the results from a set of \textit{weak learners} to produce a \textit{strong} prediction. Generally, a weak learner's performance is close to a random guess. A weak learner divides the job of a single predictor across many weak predictor functions, and optimally combines the votes from all smaller predictors. This helps enhancing the overall prediction accuracy.

XGBoost is a tool developed by utilizing these boosting principles \citep{Chen2016}. XGBoost combines a large number of regression trees with a small learning rate. As regression can be used to model classifiers: here the word \textit{regression} may refer to logistic or soft-max regression for the task of classification. XGBoost uses an ensemble of decision trees. 
We describe the detailed working principle in Section 6 (XGBoost: An Exploration of Machine Learning based Classification) contained in \citep{Supp_File}.

\subsection{Classification of Data}

As a first step, data from PHL-EC is pre-processed (the authors have tried to tackle the missing values by taking mean for continuous-valued attribute, and mode for categorical attributes). Certain attributes from the database, namely \textit{P.NameKepler, S.nameHD, S.nameHid, S.constellation, S.type, P.SPH, P.interiorESI, P.surfaceESI, P.disc.method, P.disc.year, P.maxmass, P.minmass, P.inclination} and \textit{Habmoon} were removed as these attributes do not contribute to the nature of classification of habitability of a planet. Though individual ESI values (and planetary mass) do contribute to habitability determination, because the data set directly provides total value of {\it  P.ESI} -- the global ESI of the planet, these features were neglected. Following this, classification algorithms were applied on the processed data set, where in total 49 features were used.

The pursuit of finding the appropriate classification method for any classification problem requires a lot of experimentation and analysis of the nature of the data. We performed a convex hull test to understand the nature of the data, and found that the data is not linearly separable. Hence, classifiers like SVM (Support Vector Machines), k-NN (k Nearest Neighbors) and LDA (Linear Discriminant Analysis) are not expected to perform well. All these classifiers were tried as candidates for classification; as expected from the convex hull test, they did not perform well. This motivated our choice of Tree-based classifiers and, more specifically, of the XGBoost.

After understanding the nature and separability of the data, a more suitable approach was developed. In this whole classification process, the PHL-EC data set had 3411 entries, from the data set obtained in November 2016: 24 entries belonging to class mesoplanet, 13 entries belonging to class psychroplanet, and 3374 entries belonging to class non-habitable \footnote{These are different from the numbers reported above. This is natural as discovery of exoplanets is a continuous process. Please note, as and when the catalog is updated, we update the training population as well.}. The number of items in this data set was significantly more than the older data set used to have. Hence, the artificial balancing method was modified. In the new balancing method, all 13 entries from psychroplanet class were considered in a smaller data set, and 13 random and unique entries from each of the other two classes were also considered. Thus, in this case, the number of entries in a smaller, artificially balanced data set was 39. Following this, each smaller data set was balanced in the ratio of 9:4 (training:testing) and 500 iterations of training and testing were performed on each such data set. 500 such data sets were framed for analysis. To sum it up, 2,50,000 iterations of training-testing were performed. 

\subsection{Accuracy of algorithms used to Classify Proxima~b}
\begin{figure}[htbp!]
\begin{center}
\includegraphics[width=0.6\columnwidth]{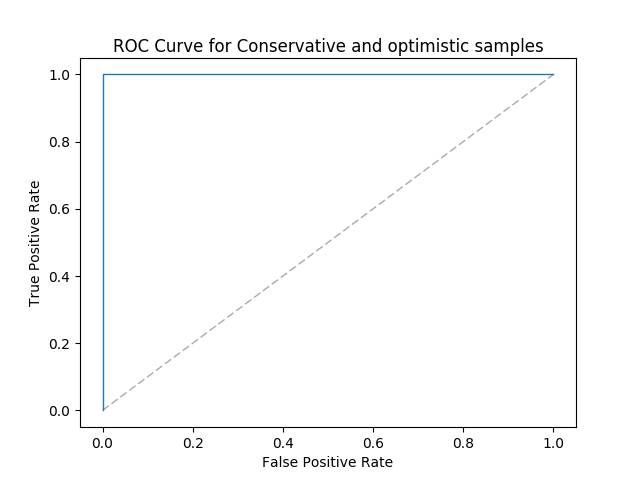}
\caption{ROC curve (blue line) for the classification of the conservative and optimistic samples: these samples include Proxima b and Kepler 186-f. 
The dotted line is the ROC for the case when the performance of the binary classifier is the same as a perfectly random guess. A good numeric representation of ROC curves is the percentage area of the coordinates which falls under the curve (AUC). Here, the AUC is $100\%$, which indicates a perfect performance of the classifier.}
\label{fig:ROC}
\end{center}
\end{figure}

XGBoost was used to classify the conservative and optimistic samples from the PHL-EC. The ROC (Receiver Operating Characteristic) curve obtained for this classification is shown in Figure~\ref{fig:ROC}. Each point on the ROC plot represents a sensitivity/specificity pair which corresponds to a particular decision threshold. Sensitivity, or recall, is the the proportion of positive tuples that are accurately identified, and specificity is the the proportion of negative tuples that are correctly identified. A test with non-overlapping classes has an ROC plot that passes through the upper left corner ($100\%$ sensitivity, $100\%$ specificity). Therefore, the closer the ROC plot is to the upper left corner, the higher the overall accuracy of the test \citep{zweig}.

The accuracy of the XGBoost algorithm used to classify Proxima~b was 100\%, i.e. there were no false positives or false negatives in its classification. The method of classification was to select Proxima~b as the training set and the remaining samples in the catalog as the test set (subject to artificial balancing by under-sampling the non-habitable class).

However, in cases where a dataset exhibits a data bias towards one class, the F-score test statistic is more representative than the accuracy of a classifier. It is used to analyze whether a classifier is able to achieve both high precision and high recall simultaneously (for details on precision, recall, F-score, ROC curves, etc., see \citep{Peres2015,Rijsbergen1979}). The values for precision and recall were calculated to gauge the goodness of the classifier. This was done but considering the psychroplanet and mesoplanet classes as the positive classes one at a time. With respect to the psychroplanet class, the calculated F-score was 0.94, precision was 0.95, and recall was 0.93. With respect to the class of mesoplanets, the F-score was 0.95, precision was 0.93, and recall was 0.97. Using the XGBoost classifier, the class-belongingness of Proxima~b to the class of psychroplanets was estimated to be $100\%$. This is indeed the true class of Proxima~b. We can thus say that XGBoost performs the classification remarkably well! 

\section{The TRAPPIST-1 system} \label{sec:trappist}

This section focuses on the application of the same algorithms and methods discussed in previous sections for the classification and CDHS computation of the planets in the TRAPPIST-1 system. 
The discovery of the TRAPPIST-1 system has caught the attention of the entire astronomy community recently \citep{Gillon2016,Walkowicz}. TRAPPIST-1 is an ultra-cool dwarf, detected by the 2MASS Sky Survey. Following this study, a series of papers were published \citep{T22017} by various researchers working on the exoplanets, and for a good reason: all seven planets in the TRAPPIST-1 system are likely  Earth-sized and rocky, with the estimated low equilibrium temperatures --- due to the exceptionally low stellar luminosity (1/1000th of the Sun), the insolation on the planets is equivalent to the insolation on the terrestrial group, thus allowing the possibility of liquid water on the surface. Three of the planets are within in the stellar habitable zone. Though all planets are most probably tidally locked with the parent star, water could still exist even on the innermost planets \citep{Gillon2016}.

We have used Cobb-Douglas habitability model to compute CDHS of TRAPPIST-1 planets, as well as classify them using K-NN classifier. The result of the experiment is shown in Table~\ref{t:b2}. We conclude from the result (a consequence of Method~1, Figure~\ref{fig:flowchart}) that the planets are in the ``Earth-League". 

\begin{table*}[ht!]
\caption{Observed and calculated parameters of TRAPPIST-1 planets. Physical parameters are given in Earth units (EU). Computed habitability score and class of the planets belonging to  planets system satisfy the threshold condition (Algorithm 3). The outcome of the machine learning algorithm fortifies the experimental findings. Please refer to Figure~\ref{fig:flowchart} for visual illustration. Class label 6 implies most likely habitable exoplanets in a probabilistic sense (see Earth-League candidate selection algorithm in Section 4 (Algorithm for Habitability candidacy classification) of \citep{Supp_File} and \citep{Bora2016CDHS}.)}
\centering
\begin{tabular}{cccccccccc}
\hline
%\toprule
Planet & Mass & Radius & Mean Insolation & Mean $T_{s} (K)$ &   CDHS$_{DRS}$ & CDHS$_{CRS}$ & Class & {\it  P.ESI} \\
   \hline
%\midrule
b & 0.86   & 1.09   &  4.2     & 396.5  &  1.0318 & 1.0410 & 5 & 0.56\\
c & 1.38    & 1.06   & 2.25   & 347.9 &    1.14084 & 1.1589 & 5 & 0.73\\
d & 0.41    & 0.77   & 1.13   & 292.4 &  0.9642 & 0.8870  & 5 & 0.9\\
e & 0.64    & 0.92    &  0.65  & 260.4 &  0.9722 & 0.9093  & 6 & 0.85\\
f & 0.67   &  1.04     & 0.38  & 229.7  &    0.9803 & 0.9826 & 6 & 0.68\\
g & 1.34    & 1.13     & 0.26   & 216.1 &  1.0951 & 1.1085 & 6 & 0.58\\
h & 0.35   &  0.75     & 0.14   & 181.8 &   0.9511 & 0.8025  & 5 & 0.45\\
    \hline
    \end{tabular}
\label{t:b2}
\end{table*}

\section{Discussion and Conclusion} \label{sec:conclusion}

The discovery of Proxima~b was announced on 24th August 2016 \citep{Anglada-Escudé}. Proxima~b is, at least, 1.3 times heavier than Earth. According to the PHL-EC, its radius is 1.12 EU, density is 0.9 EU, surface temperature is 262.1 K, and escape velocity is 1.06 EU. These attributes are close to those of the Earth, hence, there are plausible reasons to believe that Proxima~b may be a habitable planet. In the PHL-EC data set, Proxima~b is classified as a psychroplanet. We have computed the habitability score CDHS of Proxima~b using different combination of planetary  parameters:  radius, density, escape velocity and surface temperature; only surface temperature and radius; stellar flux and radius; and stellar flux and mass. According to our classification algorithm, Proxima~b falls in the Earth's class -- ``Earth-League" \citep{Bora2016CDHS}. Its habitability ``floor function value" is 1, and the difference between its CDHS and the Earth's CDHS is within the acceptable threshold of 1, as discussed in the published paper. The classification model XGBoost was used in this work to classify Proxima~b with an accuracy of 100\% (no false negatives or false positives). The accuracy results provide evidence of the strength of the model to automatically label and classify newly discovered exoplanets, such as Proxima~b, or TRAPPIST-1 planets) in this case.

\begin{table}[hbtp!]
\caption{Habitability score of Proxima~b and Kepler-186~f, two planets most potentially habitable planets before the TRAPPIST-1 system  discovery. Earth's ESI and CDHS are both 1 and considered as the baseline. CDHS values of the exoplanets in this table and Table \ref{t:b2} are significantly closer to the baseline score (Earth's score) compared to the ESI.}
\centering
\begin{tabular}{l c c c}
\hline
 Planet Name & CDHS (DRS) & CDHS (CRS) & {\it  P.ESI} \\
\hline
Kepler 186 f & 1.075074 & 1.086295 & 0.61 \\
Proxima Centauri b & 1.08297 & 1.095255 & 0.87 \\
\hline
\end{tabular}
\label{t:b3}
\end{table}

Our algorithm is emphatically exhibiting the validation of the potential habitability of Proxima~b, matching with the PHL findings. The robustness of the formula and the solid theory behind the formulation are validated by the proximity of the scores computed for  different cases. 
We have worked on two ways of affirming the habitability score of a planet. Essentially, we answer two questions: ``\textit{Is this planet potentially habitable?}" and ``\textit{How potentially habitable is this planet?}". These two questions are like the two sides of the same coin. By performing classification, we can affirm if a planet is expected to be potentially habitable or not, and by computing the CDHS, we are basically assigning a number to every planet which reflects its potential habitability. By doing this, in the future, we can gain deeper insights to planet's characteristics, understand what range of scores of the CD-HPF implies which classes of habitability, approximate unobserved attributes of a planet, etc. As the volume of data in the PHL-EC catalog increases with time, a robust automated method must be in place to analyze the data quickly and in an efficient way. An automated system primarily serves two purposes. The first is that it reduces human error in computation. The second, it eradicates the subjectivity that arises when different people try to classify or judge any data sample (here, a planet). One researcher's appraisal of a data sample might not be the same as that of another when evaluated based on general characteristics. However, when an algorithm is used for this, the results will be the same, regardless of which computer the system is deployed on. Hence, the implication of a system like this is the  standardization of classification and of multiple ways of evaluating the potential habitability of an exoplanet.

\begin{table}
\caption{Summary of results of both methods: samples which are labeled as Class 6, an indicator of potential habitability are also predicted as habitable: the outcome of both approaches matches. For example, Trappist~1-e which is labeled as psychroplanet by Method~2 with 100\% accuracy is also in Class 6 (most likely habitable class, \citep{Bora2016CDHS})  according to the classification method~1. Kepler 186 f could not be tested with a classifier as it belongs to the class of \textit{hypopsychroplanets}, which has a total of only three samples. Thus, it is unsuitable to test the classification algorithms on.}
\label{t:aug:abhijit}
\centering
\begin{tabular}{|c | c c c | c c|}
\hline

\multirow{2}{*}{Exoplanet}
& \multicolumn{3}{ c |}{Method 1: Explicit Score Calculation} & \multicolumn{2}{ c |}{Method 2: Classification by XGBoost}\\ \cline{2-6}
 & $CDHS_{DRS}$ & $CDHS_{CRS}$ & Class Category & Accuracy (\%) & Predicted Class\\ \hline

Kepler 186 f & 1.075074 & 1.086295 & 6 & -- & --\\
Proxima b & 1.08297 & 1.095255 & 6 & 100.0 & psychroplanet\\
TRAPPIST-1 b & 1.0318 & 1.0410  & 5 & 89.6 & non-habitable\\
TRAPPIST-1 c & 1.14084 & 1.1589 & 5 & 88.4 & non-habitable\\
TRAPPIST-1 d & 0.9642 & 0.8870 & 5 & 100.0 & mesoplanet\\
TRAPPIST-1 e & 0.9722 & 0.9093 & 6 & 100.0 & psychroplanet\\
TRAPPIST-1 f & 0.9803 & 0.9826 & 6 & 99.7 & psychroplanet\\
TRAPPIST-1 g & 1.0951 & 1.1085 & 6 & 82.3 & psychroplanet\\
TRAPPIST-1 h & 0.9511 & 0.8025 & 5 & 95.1 & non-habitable\\

\hline
\end{tabular}
\end{table}

This system may be extended in the future to analyze how the CDHS correlate with the classes of habitability. As there are multiple classes, it would be interesting to see if the CDHS falls into certain ranges for each class. The convergence of the score, however, gives rise to the following questions: 
\begin{enumerate}
\item Are only stellar flux/surface temperature and radius/mass enough to construct a reliable habitability score via machine learning?
\item Should a full-scale dimensionality reduction technique be employed (completely data-driven approach) in the future, to analyze the context and validate such a claim? 
\end{enumerate}

Context is critical in solving a problem as complex as determining the habitability of discovered exoplanets. We mention with great regard the advances and contributions made by Dirk Schulze-Makuch, Abel Mendez, and other researches working in this field. In contrast to the ESI metric, our approach is entirely data-driven and inspired by machine learning. Our methods and algorithm cross-match the observation that Proxima~b falls in the Earth category and is potentially habitable. This is an ample testimony of the efficacy of the proposed work.

CD-HPF is a novel metric of defining habitability score for exoplanets. It needs to be noted that the authors perceive habitability as a probabilistic measure, or a measure with varying degrees of certainty. Therefore, the construction of different classes of habitability classes 1 to 6 is contemplated, corresponding to measures as ``most likely to be habitable" as Class~6, to ``least likely to be habitable" as Class~1. As a further illustration, classes~6 and 5 seem to represent the identical patterns in habitability, but they do not! Class~6 -- the ``Earth-League" -- is different from Class~5 in the sense that it satisfies the additional conditions of thresholding and probabilistic herding and, therefore, ranks higher on the habitability score. This is in stark contrast to the binary definition of exoplanets being ``habitable or non-habitable", and a deterministic perception of the problem itself. The approach therefore required classification methods that are part of machine learning techniques and convex optimization --- a sub-domain strongly coupled with machine learning. Cobb-Douglas function and CDHS are used to determine habitability and the  maximum habitability score of all exoplanets with confirmed surface temperatures in the PHL-EC. Global maxima are calculated theoretically and algorithmically for each exoplanet, exploiting intrinsic concavity of CD-HPF and ensuring no curvature violation. Computed scores are fed to the attribute enhanced K-NN algorithm --- a novel classification method, used to classify the planets into different classes to determine how similar an exoplanet is to Earth. The authors would like to emphasize that, by using classical K-NN algorithm and not exploiting the probability of habitability criteria, the results obtained were pretty good, having 12 confirmed potentially habitable exoplanets in the ``Earth League". We have created a web page ({\tt https://habitabilitypes.wordpress.com/}) for this project to host all relevant data and results: sets, figures, animation video and a graphical abstract. The web page contains the full customized catalog of all confirmed exoplanets with class annotations and computed habitability scores. The  catalog is built with the intention of further use in designing statistical experiments for the analysis of the correlation between habitability and the abundance of elements (this work is briefly outlined in \citep{Safonova2016}). It is a very important observation that our algorithm and method give rise to a score metric, CDHS, which is structurally similar to the PHI (Planetary Habitability Index; \citep{Schulze2011Two-tired}) as a corollary in the CRS case (when the elasticities are assumed to be equal to each other). Both are geometric means of the input parameters considered for the respective models.

CD-HPF uses four parameters (radius, density, escape velocity and surface temperature) to compute habitability score, which by themselves are not sufficient to determine habitability of exoplanets. Sometimes, there is a missing data in the catalogs, such as the case with 11 rocky planets mentioned in Section~\ref{sec:analytical_cdhs}B. The unknown surface temperatures (or other parameters) can be estimated using various statistical models. In addition, parameters such as e.g. orbital period, stellar flux, distance of the  planet from host star, etc. may be equally important to determine the habitability. Future work may include incorporating more input parameters to the Cobb-Douglas function, coupled with tweaking the attribute-enhanced K-NN algorithm by checking an additional condition. Cobb-Douglas, as proved, is a scalable model and doesn't violate curvature with additional predictor variables. However, it is pertinent to check for the dominant parameters that contribute more towards the habitability score. This can be accomplished by computing percentage contributions to the response variable -- the habitability score. We would like to conclude by stressing on the efficacy of the method of using a few of the parameters rather than sweeping through a host of properties listed in the catalogs, effectively reducing the dimensionality of the problem.

To sum up, CD-HPF and CDHS turn out to be self-contained metrics for habitability. We would like pose the following questions in this context:
\begin{itemize}
\item How the two approaches coincide/converge?
\item What is the implication in the overall scientific context?
\end{itemize}
The CDHS of Kepler 186-f planet turns out to be very close to the Earth's. The habitability potential of Kepler 186-f, estimated via Earth similarity, was computed by the explicit approach (Method~1, Figure~\ref{fig:flowchart}). The implicit approach (Method~2, Figure~\ref{fig:flowchart}) assigns Kepler 186-f the label of a habitable class of exoplanets. We observe the identical scenario in the case of the TRAPPIST-1 system, where the CDHS of the TRAPPIST-1 planets align with expected class labels.

The concept of developing a classifier based on our growing knowledge of exoplanets is intriguing. There is no reason why such an approach shouldn't work, other than to think that of the large number of possible habitable exoplanets. We have parameters based on only one example that is known to be habitable and in that regard assume that all non-Earth like exoplanets are non-habitable. Our definition of habitability may need to be refined as we find more truly habitable planets.

We make use of stochastic gradient ascent to find local maxima. Evolutionary  algorithms may also be used to track dynamic functions of the type that allow for the oscillation that are instead mitigated with SGA. Additionally, we make use of 49 features through XGBoost. The results suggest that the use of Proxima b for training and remaining samples in the catalog for testing performed well with XGBoost (AUC 1.0) which is surprisingly good. We wonder a neural network may work as well. If XGBoost has AUC $> 0.99$ on training, one would expect even a vanilla feed-forward neural network trained with back propagation would have similar accuracy. Additionally, it would be interesting to try a fuzzy approach on this problem where planets have membership in all class labels but just to differing degrees. Given the sparsity of our knowledge about planets, their features, and habitability, a fuzzy approach may be worth exploring and comparative analysis with traditional classification approaches may be documented.

Rapid discoveries of exoplanets notwithstanding, it is unrealistic and premature to predict how Earth-like are the conditions on any planet on the basis of the scant data. The best case scenario is to adduce a list of optimistic targets for future detailed missions. This manuscript achieves that goal by combining physical observations, mathematical rigor and machine learning techniques. However, our approach might pay rich dividends as encouraging observations have been reported very recently, \citep{Water_Trapp}. Space Telescope Imaging Spectrograph (STIS) has been used to study the amount of ultraviolet radiation received by the TRAPPIST-1 planets. This helps determine the water content of the seven planets. The three planets within the star 's habitable zone,  TRAPPIST- 1e, f and g may possess abundant amounts of water on their surfaces indicating habitability. We predict the same using machine learning and sophisticated modeling reported in the main paper and the supplementary file, (section 5 of \citep{Supp_File}). By earth similarity approach (Method 1, Figure \ref{fig:flowchart}), we obtain the habitability scores of TRAPPIST-1e, f and g close enough to Earth (within two decimal places,Table \ref{t:aug:abhijit}). These planets are also classified as habitable by boosted tree learning (Method 2, Figure \ref{fig:flowchart}, Table \ref{t:aug:abhijit}). This is definitely encouraging.

%\bibliography{bib-exercise}
\end{document}